\documentclass[aps, prd, twocolumn, superscriptaddress, showkeys, longbibliography, groupedaddress]{revtex4}
\usepackage[T1]{fontenc}
\usepackage{subfig}
\usepackage{scalerel}
\usepackage{graphicx} 
\usepackage{dcolumn}   
\usepackage{bm}        
\usepackage{amssymb}   
\usepackage{lipsum}
\usepackage{lineno}
\usepackage{url}
\usepackage{amsmath,array}
\usepackage{natbib}
\usepackage{txfonts}
\usepackage[normalem]{ulem}
\usepackage{soul}
\usepackage{flushend}
\usepackage{hyperref}
\hypersetup{
	colorlinks=true,
	linkcolor=blue,
	filecolor=magneta,      
	urlcolor=blue,
}
\usepackage{tikz}
\usetikzlibrary{svg.path}
\definecolor{orcidlogocol}{HTML}{A6CE39}
\tikzset{
	orcidlogo/.pic={
		\fill[orcidlogocol] svg{M256,128c0,70.7-57.3,128-128,128C57.3,256,0,198.7,0,128C0,57.3,57.3,0,128,0C198.7,0,256,57.3,256,128z};
		\fill[white] svg{M86.3,186.2H70.9V79.1h15.4v48.4V186.2z}
		svg{M108.9,79.1h41.6c39.6,0,57,28.3,57,53.6c0,27.5-21.5,53.6-56.8,53.6h-41.8V79.1z M124.3,172.4h24.5c34.9,0,42.9-26.5,42.9-39.7c0-21.5-13.7-39.7-43.7-39.7h-23.7V172.4z}
		svg{M88.7,56.8c0,5.5-4.5,10.1-10.1,10.1c-5.6,0-10.1-4.6-10.1-10.1c0-5.6,4.5-10.1,10.1-10.1C84.2,46.7,88.7,51.3,88.7,56.8z};
	}
}
\newcommand\orcidicon[1]{\href{https://orcid.org/#1}{\mbox{\scalerel*{
				\begin{tikzpicture}[yscale=-1,transform shape]
				\pic{orcidlogo};
				\end{tikzpicture}
			}{|}}}}
\begin{document}
 \title{
 	Magnetic fields in a hot dense neutrino plasma and the Gravitational Waves}
\author{Arun Kumar Pandey \orcidicon{0000-0002-1334-043X}\,}
\email{arunp77@gmail.com}
\affiliation{%
	Department of Physics and Astrophysics, University of Delhi, Delhi 110 007, India}
\author{Pravin Kumar Natwariya \orcidicon{0000-0001-9072-8430}\,}
\email{pravin@prl.res.in}
\affiliation{%
    Physical Research Laboratory, Theoretical Physics Division, Ahmedabad 380 009, India}
\affiliation{%
    Department of Physics, Indian Institute of Technology, Gandhinagar 382 424, India}  
\author{Jitesh R Bhatt \orcidicon{0000-0001-7465-8292}}
\email{jeet@prl.res.in}
\affiliation{
	Physical Research Laboratory, Theoretical Physics Division, Ahmedabad 380 009, India}
\date{\today}
\begin{abstract}
    {\centering
        \bf Abstract\par}
    In the present work, we have studied the spectrum of the primordial gravitational waves due to magnetic instability in the presence of neutrino asymmetry. The magnetic instability generates a helical magnetic field on a large scale. The anisotropic stress generated by the magnetic field shown to be a source of primordial gravitational waves (GWs) at the time of matter-neutrino decoupling. We expect that the theoretically predicted GWs by this mechanism may be detected by Square Kilometer Array (SKA) or pulsar time array (PTA) observations. We also compare our findings with the results obtained by the earlier work where the effect of magnetic instability was not considered.
\end{abstract}
\maketitle
\section{Introduction}
The observation of gravitational waves (GW) by the Laser Interferometer Gravitational-Wave Observatory (LIGO) has provided first direct evidence for gravitational waves \cite{abbottetal:2016}. The events detected by LIGO involves gravitational waves from black holes binary and coalescing neutrons stars \cite{Abbott:2017gw2}. This has opened up a possibility of studying new physics by observing the Universe through gravitational waves. This remarkable success of LIGO in observing GW has provided hope of detecting the primordial gravitational waves with high sensitivity detectors \cite{Moore:2014lga}. In particular, if we will be able to detect the primordially generated GW, it would open up the window to study the early universe via gravitational waves. Primordial models of generation of the GW are studied by various cosmological and particle physics models. Primordial GW can broadly be divided into two main categories: First category consists of the gravitational waves generated during inflation or before inflation by various turbulent phases of the universe, and the second category corresponds to generation mechanism between the end of inflation and the onset of the Big Bang Nucleosynthesis (BBN). In the standard inflationary scenario, the GW background is generated due to tensor vacuum fluctuations $h_{ij}$ of the Friedman-Lemaitre-Robertson-Walker (FLRW) metric \cite{Starobinsky:1979ty, Rubakov:1982df, Giovannini:1999m, Sahni:1990tx, Khlebnikov:1997di, Guzzetti:2016bl}. In the second category consists variety of sources such as cosmological defects \cite{Vachaspati:1985tv, Figueroa:2013dh, Binetruy:2012ze}, various phase transitions \cite{Kamionkowski:1994km,Kosowsky:1992tw, Kosowsky:1994tm, Riccardo:2001an, Witten:1984we, hogan:1986cj}, turbulent phase of the cosmic fluid \cite{Kosowsky:2002tm, Dolgov:2002gn, Nicolas:2004an, Niksa:2018ofa}, primordial magnetic fields \cite{Deriagin:1987gr, Kosowsky:2002tm}, from merging of primordial black holes (PBHs) \cite{GarciaBellido:1996qt, Clesse:2016ajp} and due to instabilities in early universe \cite{Anand:2017zpg}.

In the present work, we focus on the generation of primordial GW during the neutrino decoupling. First, it should be noted that the neutrino plasma interaction has been studied by several authors: the non-linear coupling of intense neutrino flux with the collective plasma oscillations in supernova core can cause parametric instabilities in the surrounding plasma \cite{BINGHAM:1996rh, BINGHAM:1994rj, Tsytovich:1998vn, Bhatt2017}. In Refs. \cite{Silva:1999lo, Silva:2006rl, Haas:2017jc}, authors have shown that neutrino driven streaming instability can cause significant energy transfer from neutrino to the plasma. The problem of primordial GW generation during the matter-neutrino decoupling has been studied by earlier works \cite{Dolgov:2001dd, Dolgov:2002gn}. In reference \cite{Dolgov:2001dd}, the authors have shown that  inhomogeneous cosmological lepton number generated by some particle physics mechanism can produce turbulence around the time when the neutrino entered the free streaming regime. This could in turn give rise to the primordial magnetic fields and the gravitational waves. This requires a net lepton number density, $N_a(x)=n_{\nu_a}(x)-n_{\bar{\nu}_a}(x)$ where $a=e,\mu, \tau$, of one or more neutrino species over some characteristic length scale which is  regarded as much smaller than the Hubble scale. A net flux of neutrinos along the gradient then might chaotically stir the plasma to produce magnetic fields and the GWs.
This work was further extended in Ref. \cite{Dolgov:2002gn} and shown that the continuous energy injection to the plasma can lead to the modification of the Kolmogorov spectrum and may influence the spectrum of gravitational waves.

It should be emphasized here that it is {\it not crucial} to have the inhomogeneity in neutrino density in order to generate a magnetic field. It was pointed out in Ref. \cite{Dvornikov:2013bca} that the parity violation in the Standard Model can lead to modification of the equations of magneto-hydrodynamics (MHD) \cite{Haas:2017jc}. The modified MHD gives an anti-symmetric contribution to the photon polarization tensor arising due to parity-violating neutrino interaction with charged leptons. This is related with the Chern-Simon term in the effective Lagrangian. This, in turn, leads to a new instability of magnetic fields in the (electroweak) plasma in the presence of a nonzero neutrino left-right asymmetries. In this work, neutrino density can be homogeneous and it can produce a strong magnetic field \cite{Dvornikov:2013bca, Bhatt:2016hyi}. In the present work, we aim to study the spectrum of primordial gravitational waves generated by the instability of the magnetic modes in the presence of homogeneous neutrino asymmetry distribution. It ought to be noted here that the similar kind of instability can also occur due to non-linear interactions of neutrinos and the collective modes of the plasma \cite{Yamamoto:2016ny}. 

This paper is organized as follows: In section \eqref{sec-2},  we have given a brief overview of the effective Lagrangian in the presence of  charged lepton-neutrino interaction. In this section, we have also given a brief description of the generation of magnetic fields in subsection \eqref{sub-2.2}, and it's a contribution to the energy-momentum tensor in subsection \eqref{sub-2.3}. Section \eqref{sec-3} contains the generation of GW due to magnetic fields produced in an asymmetric neutrino plasma. In this section, we have derived the evolution equation of the GW and derived the formula for the current power spectrum. Finally, we have summarized and discussed the results obtained in section \eqref{sec-4}.
Throughout this work, we have used natural units and the background space-time line element, given by Friedman-Robertson-Walker metric
\begin{equation}
ds^2=a^2(\tau)\left(-d\tau^2+\delta_{ij}\,dx^i\,dx^j \right)\, ,
\label{eq:FLRW-backGround}
\end{equation}
where scale factor $a(\tau)$ have dimension of length, whereas conformal time $\tau$ and conformal coordinate $x^i$ are dimensionless quantities. In the radiation dominated epoch $a=1/T$, we can define conformal time $\tau = M_*/T$, where $M_* = \left({90\over 8\pi^3g_{{\rm eff}}}\right)^{1/2}M_{pl}\,, $ $g_{{\rm eff}}$ and $M_{pl}=1/\sqrt{G}$ are effective relativistic degree of freedom and Planck mass respectively. Also, we have used comoving variables defined as ${\bf B}_c({\bf x})= {\bf B}({\bf x}, \tau)\, a^2(\tau)$ (magnetic field), $\mu_c =a\, \mu$ (chemical potential) and $\sigma_c= a\, \sigma $ (electrical conductivity). However, in the upcoming sections, we will not use the subscript `c' for the simplicity of the notation. 
\section{Neutrinos in hot dense plasma medium}\label{sec-2}
Neutrinos behave differently in a dense matter than in vacuum, which could be very useful to understand the neutrino oscillations in the early universe \cite{Bahcall:2004ut, Fuller:1987rn, Dolgov:1980cq}. In the present work, we have considered a massless neutrino in a hot plasma of charged electrons. In this case, the evolution equation of a charged lepton $l$, represented by a bispinor  $\psi$ in a neutrino-antineutrino gas (i.e. $\nu_\alpha\, \bar{\nu}_\alpha$-gas, here $\alpha=e, \mu, \tau$) is given by the following Dirac equation \cite{Studenikin:2008qk, Giunti:2007ry}
\begin{equation}
[i\gamma^\rho \partial_\rho-\gamma_\rho(f^\rho_L\, P_L+f^\rho_R\, P_R)]\psi =0,
\end{equation}
here, $\gamma^\rho=(\gamma^0, \gamma^i)$ and $P_{L,R}=(1\mp\gamma^5)/2$ are the Dirac matrices and the chiral projection operators respectively, $\gamma^5=i\,\gamma^0 \gamma^1 \gamma^2 \gamma^3$. The external neutrino macroscopic currents $f^{\rho}_{L,R}=(f^0_{L,R}, \textbf{f}_{L,R})$ describes the $l\nu$ interaction in the mean field approximation. The `$\nu l$-interaction' is given by the following effective Lagrangian \cite{Giunti:2007ry}:
\begin{eqnarray}
\mathcal{L}_{\rm eff}= -\sqrt{2} G_F \sum_\alpha \, [\bar{\nu}_\alpha \gamma^\rho (1-\gamma^5)\nu_\alpha]\, f_\rho,
\end{eqnarray}
where, $G_F\approx 1.17\times 10^{-5}$ GeV$^{-2}$ is the Fermi constant. Here $f_\rho$ is defined as
\begin{equation}
f_\rho= [\bar{\psi}\, \gamma_\rho a_L^{(\alpha)}P_L\, \psi+ \bar{\psi}\,\gamma_\rho a_R^{(\alpha)}\, P_R\, \psi],
\end{equation}
The coefficients $a_{L,R}^{(\alpha)}$ are the given in terms of Weinberg angle $\theta_W$ by (for details, see \cite{Giunti:2007ry}, page number 78 and 138)
\begin{equation}
a_{L}^{(\alpha)}  = -\frac{1}{2}+\sin^2 \theta_W +\delta_{\alpha, e} ~~~~~~~~~
a_{R}^{(\alpha)}  = \sin^2 \theta_W.
\end{equation}
Here, $\delta_{e,e}=1$ else $\delta_{l \neq e ,e}=0$. Average over the neutrino ensemble of the effective Lagrangian  $\langle \bar{\nu}_\alpha \gamma^0 (1-\gamma^5)\nu_\rho\rangle \approx 2\Delta n_{\nu_\alpha} $, which is zeroth component of the macroscopic neutrino current. Here $\Delta n_{\nu_\alpha}=n_{\nu_\alpha}-n_{\bar{\nu}_\alpha}$ is the number asymmetry of the neutrinos and antineutrinos at temperature  $T_{\nu_\alpha}$ and
\begin{equation}
n_{\nu_\alpha, \bar{\nu}_\alpha}= \int \frac{d^3 p}{(2\pi)^3}\frac{1}{1+\exp\left(\frac{|{\bf p}|\mp \mu_{\nu_\alpha}}{T_{\nu_\alpha}}\right)}\, ,
\end{equation}
In the above equation, we assume that the chemical potential of the neutrinos and antineutrinos are equal i.e., $|\,\mu_{\nu_\alpha}|=|\,\mu_{\bar{\nu}_\alpha}|$. It is shown in reference \cite{Dvornikov:2013bca} that $\textbf{f}_{L,R}=0$ and 
\begin{eqnarray}
f_L^0 & = &2\sqrt{2} \, G_F \left[\Delta n_{\nu_e}+\left(\sin^2\theta_W -\frac{1}{2}\right)\sum_\alpha \Delta n_{\nu_\alpha}\right], \\
f_R^0 & = & 2\sqrt{2} \, G_F \sin^2\theta_W \sum_\alpha \Delta n_{\nu_\alpha}\,.
\end{eqnarray}
The difference between these two currents:
\begin{eqnarray}
f_L^0 -f_R^0 & = & 2\sqrt{2} G_F \, [\Delta n_{\nu_e}-\frac{1}{2}\sum_\alpha \Delta  n_{\nu_\alpha}], \\
& = & \sqrt{2} G_F \, [\Delta n_{\nu_e}-\Delta n_{\nu_\mu}-\Delta n_{\nu_\tau}]\,, \\
& = & \frac{\sqrt{2} G_F\, T^3}{6} \,[\xi_{\nu_e}-\xi_{\nu_\mu}-\xi_{\nu_\tau}].
\end{eqnarray}
Above we have used a relation for number asymmetry of a neutrino species as $\Delta n_{\nu_\alpha} =\xi_{\nu_\alpha} T^3/6$, where $\xi_{\nu_\alpha} = \mu_{\nu_\alpha}/T$. 
\subsection{Current expression for asymmetric neutrino gas}\label{sub-2.1}
The general form of the Maxwell equation is given by the equation  $\partial_\nu\, F^{\mu\nu} =j^\mu$, where $j^\mu$ is the total current $j^\mu= j^\mu_{\rm ext}+ j^{\mu}_{\rm neu}$.  The current $j^{\mu}_{\rm neu}$ is the neutrino current in the $\nu\bar{\nu}-$gas. $j^\mu_{\rm ext}$ is the external current. $j^{\mu}_{\rm neu}$ can be written as: $j^{\mu}_{\rm neu}=\Sigma^{\mu\nu} ({\bf K})\, A_\nu ({\bf K})$, where $\Sigma_{\mu\nu}$ is the (retarded) self-energy of the photons and it can be split in three component as
\begin{equation}
\Sigma_{\mu\nu}=P_{\mu\nu}^L\Sigma_L +P_{\mu\nu}^T\, \Sigma_T +P_{\mu\nu\beta}^A\, (f^\beta_L-f^\beta_R)\,\Sigma_A\,,
\end{equation}
where, $P_{\mu\nu}^L=\frac{k_\mu k_\nu}{k^2}$, $P_{\mu\nu}^T=\left(g_{\mu\nu}-\frac{k_\mu k_\nu}{k^2}\right)$ and $P_{\mu\nu\beta}^A=i\epsilon_{\mu\nu\alpha\beta} \, k^\alpha$. Here $\Sigma_L$, $\Sigma_T$ and $\Sigma_A$ are the form factor of the photons. The form factor $\Sigma_A$ corresponds to the parity violating part of the polarization tensor for the gas. The parity violating form factor comes because of the the neutrino-neutrino interactions and the neutrino-lepton interactions, i. e. $\Sigma_2=(f^0_L-f^0_R)\Sigma_A$ is coming from the neutrino-neutrino interactions and the neutrino-lepton interactions, i. e.
\begin{eqnarray}
\Sigma_2=\Sigma^{(\nu)}_2+\Sigma^{(\nu\, l)}_2 \, .
\end{eqnarray}
For a low density $\nu\bar{\nu}$-gas as a background for the electrons \cite{Dvornikov:2013bca},
\begin{equation} \label{eq:pi2}
\Sigma^{(\nu)}_2 =-\frac{2}{3}(f^0_L-f^0_R)\frac{\alpha_{\rm em}^2\, n_e}{m_e^3},
\Sigma^{(\nu l)}_2 = - \frac{7\pi}{3}(f^0_L-f^0_R)\, \frac{\alpha_{\rm em}\, n_e}{m_e^3}.
\end{equation}
These relation can be written symbolically as: $\Sigma_2=\frac{\alpha_{\rm em}}{\pi}(f^0_L-f^0_R)\,  \mathcal{F}$, where $\mathcal{F}$ is a dimensionless function and for a relativistic plasma $\mathcal{F}=-0.18$ and  for a relativistic degenerate plasma $\mathcal{F}=-2.05$ \cite{Dvornikov:2013bca}.
\subsection{Neutrino asymmetry and the magnetic fields}\label{sub-2.2}
The total three current in the present case can be written as
\begin{equation}
{\bf j} = \sigma \, ({\bf E+{\bf v}\times{\bf B}})+\Sigma_2\, {\bf B}+\Sigma_\omega \boldsymbol{\omega} \label{eq:current2}\,.
\end{equation}
Here, $\boldsymbol{\omega}$ is the vorticity three vector. In the above equation, the first term is the conduction current. The second and third term comes only for the neutrino asymmetric $\nu\bar{\nu}-$gas. Using eq. \eqref{eq:current2} and Maxwell's equation, evolution equation for the magnetic field can be written as:
\begin{equation}
\frac{\partial {\bf B}}{\partial \tau}=\frac{\nabla^2 {\bf B}}{\sigma}+\nabla\times({\bf v}\times {\bf B})+\frac{\Sigma_2}{\sigma}\nabla \times {\bf B} +\frac{\Sigma_\omega}{\sigma} \nabla\,\times \boldsymbol{\omega}.
\label{eq:induc1}
\end{equation}
The first term on the right-hand side is the diffusion term; the second term is the convection term, and the remaining two terms are due to the neutrino asymmetry. The last term in the above equation comes due to the non-conversion of the total helicity 
\cite{Yamamoto:2016ny}.  For $\mu\ll T$, $\Sigma_\omega = T^2/24$ and for $\mu \gg T$, $\Sigma_\omega =-\Delta \mu^2/24\pi^2$ (here $\Delta \mu=\mu_{\nu_\alpha}^R-\mu_{\nu_\alpha}^L$ and $\Delta \mu^2=\Delta \mu(\mu_{\nu_\alpha}^R+\mu_{\nu_\alpha}^{L})$. For a causal process, coherence length scale of the magnetic fields $l_B$ should be less than the Horizon size, i.e. $l_B\sim 1/|\Sigma_2|< l_H\sim H^{-1}=M_*/T^2$ (here $\Sigma_2=\frac{\alpha_{\rm em}}{\pi}(f^0_L-f^0_R)\,  \mathcal{F}$ and $\mathcal{F}=-0.18$. For detail see the equation \eqref{eq:pi2} and reference \cite{Dvornikov:2013bca}). Thus, using this relation, we will have for $T\gg \mu_{\nu_\alpha}$ (ultra relativistic case) \cite{Dolgov:2001dd, Dvornikov:2013bca},
\begin{equation}
[\xi_{\nu_e}-\xi_{\nu_\mu}-\xi_{\nu_\tau} ]>\frac{1.1\times 10^{-6}\sqrt{g_{\rm eff}/106.75}}{(T/MeV)}\,,
\end{equation}
here, number 106.75 correspond to the relativistic effective degree of freedom at $T> T_{\rm QCD}\simeq 150$ MeV. However for the case when $T< \mu_{\nu_\alpha}$, one can calculate this limit in the similar fashion and this will be important at the core of a  supernova, where $\mu_{\nu_\alpha}\sim 100$ MeV and $T\sim 10$ MeV \cite{Yamamoto:2016ny}. In the present work, we are interested in the regime $T>1$~GeV and therefore, we consider only the case $T\gg \mu$.

Initially, when there are no magnetic fields, only last term survives, and thus it acts as a source for the generation of magnetic fields. The collective behavior of the magnetic fields in the gas depends on the competition between these two terms. The magnitude of each term can be compared to see the dominance of a term over others. This will give an insight into the dynamics of the magnetic fields at different length scale. Let us suppose that $l_B$ is a typical coherence length scale of the magnetic fields and $\mu_{\nu_e}\sim \mu_{\nu_\mu}\sim \mu_{\nu_\tau}\sim T$, $\sigma\sim T/e^2$, $|\nabla^2 {\bf B}/\sigma| \sim B/(\sigma l_B)$, $|\nabla\times({\bf v}\times {\bf B})|\sim v B/l_B$, $|\Sigma_2( \nabla\times {\bf B})/\sigma| \sim  \Sigma_2 B/\sigma l_B$ and $|\Sigma_\omega (\nabla\times \boldsymbol{\omega})/\sigma|\sim _\omega v/l_B^2 \sigma$. Now for small velocity $v \ll \text{max}\{\frac{e^2}{l_B T}, \frac{e^2 \Sigma_2}{T}\}$ or $B\ll \frac{e^2 \Sigma_\omega}{l_B\, T}$, the convection term can be neglected. However, for $l_B \gg \min \{\frac{1}{\Sigma_2}, \frac{e^2}{v T}\}$, or when $B\ll \Sigma_\omega v$, the diffusion can be ignored. In the present work, we ignore convection term.

Now consider propagation vector in $z-$direction and magnetic and velocity fields are in plane perpendicular to the propagation vector. We can decompose equation \eqref{eq:induc1} in the polarization modes $\varepsilon_i^\pm$
\begin{equation}
\varepsilon_i^\pm =\frac{1}{\sqrt{2}} [{\bf e}_1 \pm i{\bf e}_2]\, \exp(i {\bf k}\cdot {\bf x}).
\end{equation}
Polarization modes form an orthonormal triad system of unit vectors (${\bf e}_1$, ${\bf e}_2$, ${\bf e}_3={\bf k}/k$). Here $k=|{\bf k}|$ is the magnitude of the comoving momentum. These polarization modes satisfy $\nabla\cdot \varepsilon_i^\pm=0$ and $\nabla \times \varepsilon^\pm= \pm k\, {\bf \varepsilon}$ and $\varepsilon^\pm (-{\bf k})= \varepsilon^\pm ({\bf k})$.  The velocity and the magnetic fields can be decomposed as:
\begin{eqnarray}
{\bf v}({\bf x}, \tau) & = & \sum_\lambda \int \frac{d^3 k}{(2\pi)^3} \tilde{v}^\lambda ({\bf k}, \tau)\varepsilon^\lambda (k)\,,\, \\
{\bf B}({\bf x}, \tau) & = & \sum_\lambda \int \frac{d^3 k}{(2\pi)^3} \tilde{B}^\lambda ({\bf k}, \tau)\varepsilon^\lambda (k)\,,
\end{eqnarray}
Here, $\lambda =+, -$ correspond to the two polarization. The four vector $k^\mu$, with a dimension of (length)$^{-1}$ is $k^\mu=(\omega, {\bf k})$. Here, $d^3k= |{\bf k}|^2 \, d|{\bf k}|\, d\Omega$. Therefore, equation \eqref{eq:induc1} in the component form is
\begin{eqnarray}
\frac{\partial \tilde{B}^\pm}{\partial \tau} & = & -\frac{k^2}{\sigma}\tilde{B}^\pm\, \pm\, \frac{\Sigma_2 k}{\sigma} \tilde{B}^\pm +\frac{\Sigma_\omega k^2}{\sigma} \tilde{v}^\pm\, \label{eq:tildeB}.
\end{eqnarray}
In the absence of any seed magnetic field, only last term in the above equations will survive. When a sufficiently large magnetic field is generated, first and second terms will dominate and we can write a solution of the above equation as: $\tilde{B}^\pm= \tilde{B} _0^\pm \exp[-\int \left(\frac{k^2}{\sigma} \pm \frac{\Sigma_2}{\sigma} k\right) d\tau]$. Hence $\tilde{B}^+$ grow exponentially when wave number of the mode satisfy $k <\Sigma_2$. However, $\tilde{B}^-$ will always damps exponentially. Now we multiply  $\tilde{B}^{+*}$ with the equation for $\tilde{B}^+$ and $\tilde{B}^{-*}$ with the equation for $\tilde{B}^-$ and after taking ensemble average,
\begin{equation}
\frac{\partial |\tilde{B}^\pm|^2}{\partial \tau} = 2 \left(-\frac{k^2}{\sigma}\pm \frac{\Sigma_2 k}{\sigma} \right) |\tilde{B}^\pm|^2  +\frac{\Sigma_\omega k^2}{\sigma} \langle \tilde{B}^\pm\, \tilde{v}^\pm\,\rangle  \label{eq:tildeB1++}.
\end{equation}
In the last term, we need to calculate $\langle \tilde{B}^\pm\, \tilde{v}^\pm\,\rangle $. Initially, when there is no magnetic fields, solution of the equation (\ref{eq:tildeB}) is $\tilde{B}^\pm ({\bf k}, \tau')=\int_{\tau_*}^\tau d\tau' \frac{\Sigma_\omega (\tau') k^2}{\sigma}  \tilde{v}^\pm ({\bf k}, \tau')$. Therefore, we will have
\begin{equation*}
 \langle \tilde{B}^{\pm\, *}({\bf k},\tau)\, \tilde{v}^\pm ({\bf k}',\tau)\,\rangle =\int_{\tau_*}^\tau d\tau' \frac{\Sigma_\omega (\tau') k^2}{\sigma} \langle  \tilde{v}^{\pm*} ({\bf k},\tau') \tilde{v}^\pm ({\bf k}',\tau)\rangle 
\end{equation*}
 Here $\tau_*$ is the time at which asymmetry is generated. To calculate the two point correlators of the velocity fields, we assume that the velocity fields are correlated on the eddy turnover time and uncorrelated above the eddy turnover time. Which means that $\langle  \tilde{v}^{\pm*} ({\bf k},\tau') \tilde{v}^\pm ({\bf k}',\tau)\rangle = \langle \tilde{v}^\pm ({\bf k},\tau)^2\rangle (2\pi)^3 \delta^3 ({\bf k}- {\bf k}')$ for $\tau-\tau' <2\pi/(k v)$ and  $\langle  \tilde{v}^{\pm*} ({\bf k},\tau') \tilde{v}^\pm ({\bf k}',\tau)\rangle = 0$ for $\tau-\tau' >2\pi/(k v({\bf k}, \tau))$. Here in $v({\bf k}, \tau)$ is the velocity field at length scale $l=2\pi/k$ at time $\tau$. Using the two point correlation of the velocity fields, we can write the two point correlation $\langle \tilde{B}^{\pm\, *}\, \tilde{v}^\pm\,\rangle$ as:
\begin{equation*}
 \langle \tilde{B}^{\pm\, *}({\bf k},\tau)\, \tilde{v}^\pm ({\bf k}',\tau)\,\rangle =\frac{\Sigma_\omega k^2}{\sigma}\, f({\bf k}, \tau) |v^\pm|^2 (2\pi)^3 \delta^3 ({\bf k}- {\bf k}') 
\end{equation*}
Here, $f({\bf k}, \tau)= S\frac{2\pi}{k v({\bf k}, \tau)} \, \tanh\left(\frac{k v({\bf k}, \tau)}{ 2\pi S}(\tau-\tau_*)\right)$ and $S$ is known as fudge factor. This function is a smoothing function with the property of $f({\bf k}, \tau)=\tau-\tau_*$ for $\tau - \tau_* <2\pi/(kv)$ and $f({\bf k}, \tau)=2\pi/(kv)$ for $\tau - \tau_* >2\pi/(kv)$ \cite{Tashiro:2012mf}. Using the above relation in equation \eqref{eq:tildeB1++} gives (assuming $\Sigma_2$, $\Sigma_\omega$ and $\sigma$ are constant)
\begin{equation}
\frac{\partial |\tilde{B}^\pm|^2}{\partial \tau} = 2 \left(-\frac{k^2}{\sigma}\pm \frac{\Sigma_2 k}{\sigma} \right) |\tilde{B}^\pm|^2  +\frac{2 \Sigma_\omega^2 k^4}{\sigma^2}  \, f({\bf k}, \tau)\, |\tilde{v}^\pm|^2\,
\label{eq:tildeB2++}.
\end{equation}
We have used a power law power spectrum of the velocity field $|\tilde{v}^\pm|$, satisfying $|\tilde{v}^+|=\tilde{v}^-= \pi k^{-3/2} v$. Here $\frac{1}{2}v^2=\rho_v =v_i^2 (\tau)\left(\frac{k}{k_i}\right)^n$ and $k_i$ denotes the inertial wave number and $v_i$ is the fluid velocity at this scale. For $k>k_i$, $n=-2/3$ corresponds to the Kolmogorov spectrum, however,  $k<k_i$ belongs to white noise spectrum for which $n=3$ \cite{Anand:2017zpg}.

Since we are interested in a statistically homogeneous and isotropic magnetic fields, the two point correlation of the magnetic fields i.e. $\langle {\bf B}_i({\bf r}_1) {\bf B}_j ({\bf r}_2) \rangle= C_{ij}({\bf r}_1, {\bf r}_2)$ is a function of ${\bf r}=|{\bf r}_1-{\bf r}_2|$ only. 
In terms of the Fourier amplitude of the magnetic fields, the two-point correlation of the divergence-free magnetic fields gives
\begin{equation}
\langle {\bf B}_i({\bf k}) {\bf B}_j ({\bf k}') \rangle =\frac{(2\pi)^3}{2} \, \delta({\bf k}-{\bf k}')\left[P_{ij}\, S(k)\,+\,i\, \varepsilon_{ijl}\, \hat{k}_l\, \mathcal{A}(k)\right]\,,
\end{equation}
here, $P_{ij}$ is the transverse projection operators $P_{ij}(\hat{{\bf k}})= \delta_{ij}-\hat{{\bf k}}_i\hat{{\bf k}}_j$ and $\hat{{\bf k}}_i\equiv k_i/ |{\bf k}|$, and follows $P_{ij}P_{jk}=P_{ik}$ and $P_{ij} \hat{k}_j=0$. $\varepsilon_{ijl}$ is totally antisymmetric tensor. $S(k)$ and $\mathcal{A}(k)$ represents the symmetric and antisymmetric part of the correlator. These two functions are usually referred as ``magnetic'' and ``helical'' power spectrum respectively. The magnetic energy $\rho_B$ and helical energy $\mathcal{H}_B$ can be defined using these two functions as $\rho_B(k)=2\pi k^3\, S(k)$ and $\mathcal{H}_B (k)=4\pi k^2 \mathcal{A}(k)$ and defined as
\begin{eqnarray}
\rho_B(k) & = & \frac{k^3}{4\pi^2 V} \,[|\tilde{B}^+_k|^2+|\tilde{B}_k^-|^2 ]\,,\\
\mathcal{H}_B (k) & = & \frac{k^2}{2\pi^2 V}\,[|\tilde{B}^+_k|^2- |\tilde{B}_k^-|^2 ]\,.
\end{eqnarray}
Using the definition of the magnetic energy and the helicity densities, we can write evolution equation for the magnetic energy and helicity as
\begin{eqnarray}
\frac{\partial \rho_B}{\partial \tau} & = & -\frac{2\, k^2}{\sigma}  \rho_B+ \frac{\Sigma_2 k^2}{\sigma}\mathcal{H}_B  +\frac{2 \Sigma_\omega^2 k^4}{\sigma^2} f(\tau) \rho_v \label{eq:EB-bv}\,, \\
\frac{\partial \mathcal{H}_B}{\partial \tau} & = & -\frac{2\, k^2}{\sigma}\mathcal{H}_B +\frac{4\, \Sigma_2}{\sigma} \rho_B +\frac{2 \Sigma_\omega^2 k^4}{\sigma^2} f(\tau) \mathcal{H}_v \label{eq:HB-bv}.
\end{eqnarray}
Here, $\rho_v$ and $\mathcal{H}_v$ are fluid kinetic energy and fluid helicity. For a maximally helical magnetic field, $\mathcal{H}_B$ and $\rho_B$ satisfies $|\mathcal{H}_B(k, \tau)|\leq \frac{2}{k}\, \rho_B(k, \tau)$ and this occurs when either of positive or negative magnetic modes vanishes \cite{Jedamzik:2004rb}. In such a situation, equations \eqref{eq:EB-bv} -- \eqref{eq:HB-bv} gives
\begin{equation}
\frac{\partial \mathcal{H}_B (k, \tau)}{\partial \tau} = \left( -\frac{2 k^2}{\sigma} + \frac{2 \Sigma_2 k}{\sigma}\right) \mathcal{H}_B(k, \tau)+\frac{2 \Sigma_\omega^2 k^4}{\sigma^2} f(\tau) \rho_v (k, \tau)\,. \label{eq:EB-bv2}
\end{equation}
Therefore, to see the evolution of the generated fully helical magnetic fields after it's generation, we need to solve this equations with appropriate parameters and the boundary conditions.
\subsection{Energy momentum tensor spectrum}\label{sub-2.3}
The energy momentum tensor in presence of the magnetic fields and the turbulence can be given as
\begin{eqnarray}
T_{ij} =\left[(\epsilon+p)\, u_i\, u_j +p\, \delta_{ij}\right] - \left(B_i\, B_j -\frac{1}{2}\delta_{ij}B^2\right),
\end{eqnarray}
where $u_i$, $p$ and $\epsilon$ are the fluid velocity, pressure, and the energy density respectively. The turbulence lasts for a short time as compared to the evolution of the universe and acts incoherently to generate the GWs. However, the magnetic fields behave as a coherent source and can act up to the time of matter and radiation equality \cite{Caprini:2001cdr}. Therefore, in the present work, we only consider the generation of the GWs due to the magnetic fields. The source of gravitational waves is the stress tensor of the magnetic fields and is written in Fourier space as:
\begin{eqnarray}
T_{ij}({\bf k}) & = & -\frac{1}{2(2\pi)^4\, a^{-2}}\int d^3k' [B_i({\bf k}') \, B_j({\bf k}'-{\bf k}) \nonumber\\
& - & \frac{1}{2}\delta_{ij}\, B_m({\bf k}')B_m({\bf k}'-{\bf k})],\label{eq:FLRWperturb}
\end{eqnarray}
Here we have neglected contribution from the modes outside the horizon ($ k\tau\ll 1$). Now we need to calculate the TT part of the anisotropic stress tensor, which is given by:
\begin{eqnarray}
\Pi^{\rm TT}_{ij} ({\bf k}) & = &  \Lambda_{ikjl}(\hat{{\bf k}})\,T_{kl}({\bf k}) \nonumber \\
& = &  \left(P_{ik}(\hat{\bf k})\, P_{jl}(\hat{\bf k})-\frac{1}{2}P_{ij}(\hat{\bf k})P_{kl}(\hat{\bf k})\right) T_{kl} ({\bf k}).
\label{eq:aniso-ener-mom}
\end{eqnarray}
The projection operator $\Lambda_{ikjl}(\hat{{\bf k}})$ projects onto the transverse traceless part of the stress tensor. The anisotropic stress power spectrum tensor is defined as \cite{Caprini:2001cdr, Anand:2018mgf}
\begin{equation} \label{eq:Pij-corre}
\langle \Pi^{\rm TT}_{ij}({\bf k})\Pi^{\rm *TT}_{lm}({\bf k}')\rangle 
=\frac{1}{4a^4}[M_{ijlm}f(k)+i A_{ijlm} g(k)]\, \delta({\bf k}-{\bf k}'),
\end{equation}
$M_{ijlm} $ and $A_{ijlm}$ are defined as
\begin{eqnarray}
M_{ijlm} &=& P_{il}P_{jm}+P_{im}P_{jl}-P_{ij}P_{lm}  \,,\\
A_{ijlm} &=& \frac{1}{2}\hat{{\bf k}}_k \left(P_{jm}\varepsilon_{ilk}+P_{il}\varepsilon_{jmk}+P_{im}\varepsilon_{jlk}+P_{jl}\varepsilon_{imk}\right)\,, \nonumber
\end{eqnarray}
where, $f(k)$ and $g(k)$are defined as
\begin{eqnarray}
f(k) & = & \frac{1}{4}\frac{1}{(4\pi)^2} \int d^3p\,[(1+\gamma^2)(1+\beta^2)\,S(p)S(k-p) \nonumber \\
& + & 4\,\gamma\,\beta\,\mathcal{A}(p)\mathcal{A}(k-p) \, ]\, ,\\
g(k) & = & \frac{1}{2}\frac{1}{(4\pi)^2}\int d^3p\, \left[(1+\gamma^2)\,\beta\,S(p)\, \mathcal{A}(k-p)\right]\,,
\end{eqnarray}
where, $\gamma= \hat{{\bf k}}\cdot \hat{{\bf p}}$ and $\beta= \hat{{\bf k}}\cdot(\widehat{{\bf k}-{\bf p}})= (k-p\gamma)/\sqrt{k^2+p^2-2\gamma p k}\,$.
\begin{figure*}[ht]
	\centering
	\subfloat[]{\includegraphics[width=3.1in,height=2.2in]{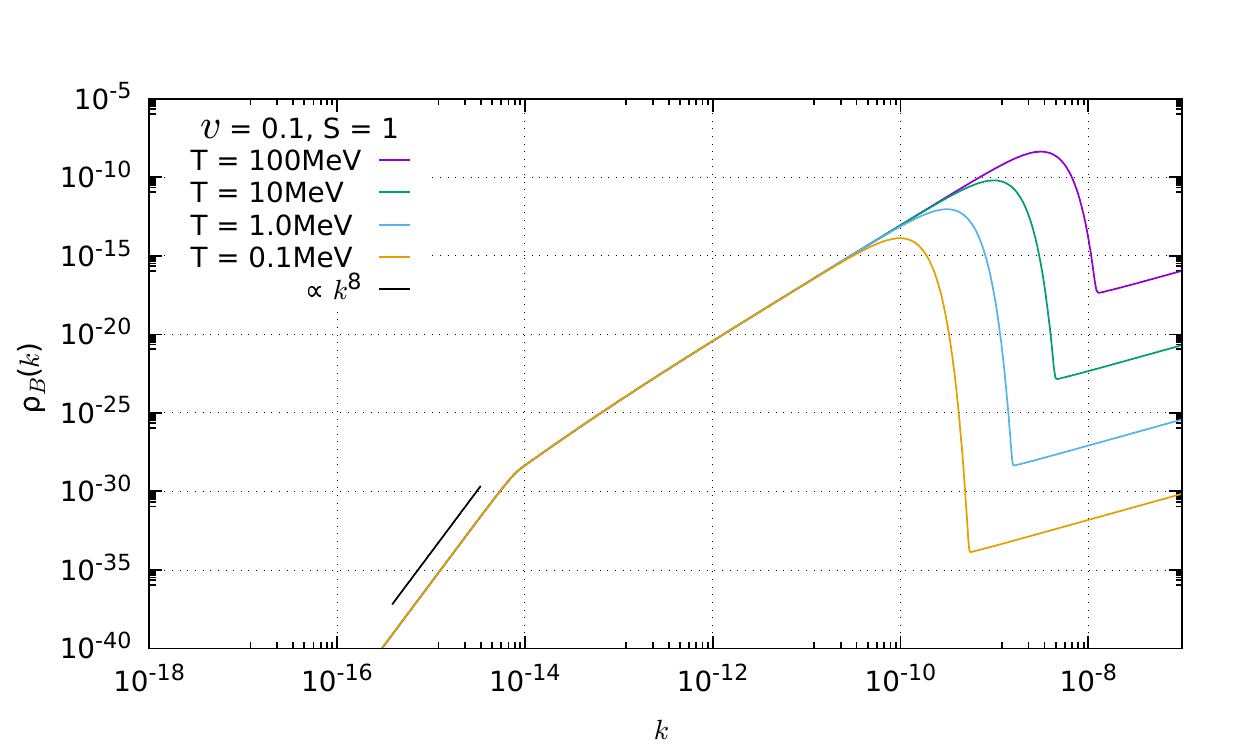}\label{fig:Eb-plot}}
	\subfloat[]{\includegraphics[width=3.1in,height=2.2in]{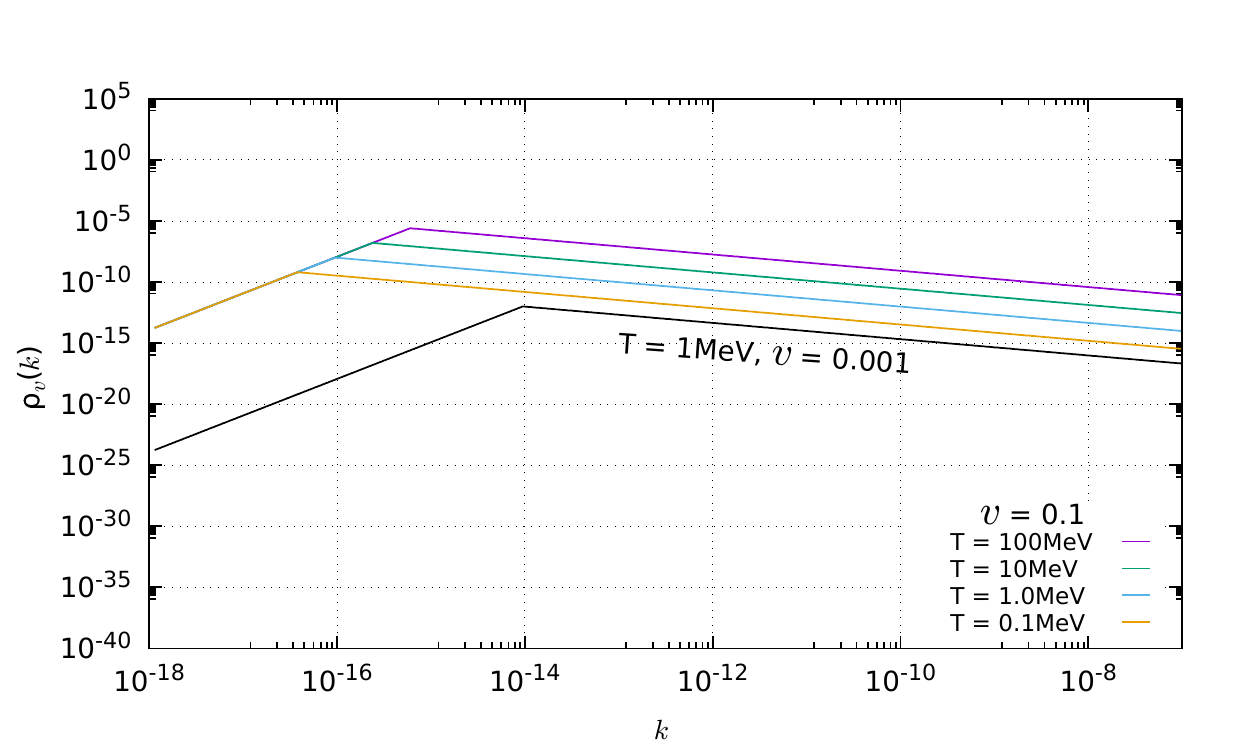}\label{fig:Ev-plot}} 
	\caption{ Plot of magnetic energy and fluid velocity spectrum with respect to wave number $k$ at different temperatures. Here we have assumed $\sigma=70$. In figure (\ref{fig:Eb-plot}), we have plotted spectrum of a maximally helical magnetic field for $v_i(\tau_*)=10^{-1}$, S = 1. Figure  (\ref{fig:Ev-plot}), correspondence to the kinetic energy of the turbulent plasma, for two different $v_i(\tau_*)=10^{-1},10^{-3}$. Here for large $k$, we have considered a Kolmogorov Spectrum and for smaller values of $k$, it's a white noise spectrum.}
	\label{fig:detail}
\end{figure*}
\begin{figure*}[ht]
	\centering
	\subfloat[]{\includegraphics[width=3.1in,height=2.2in]{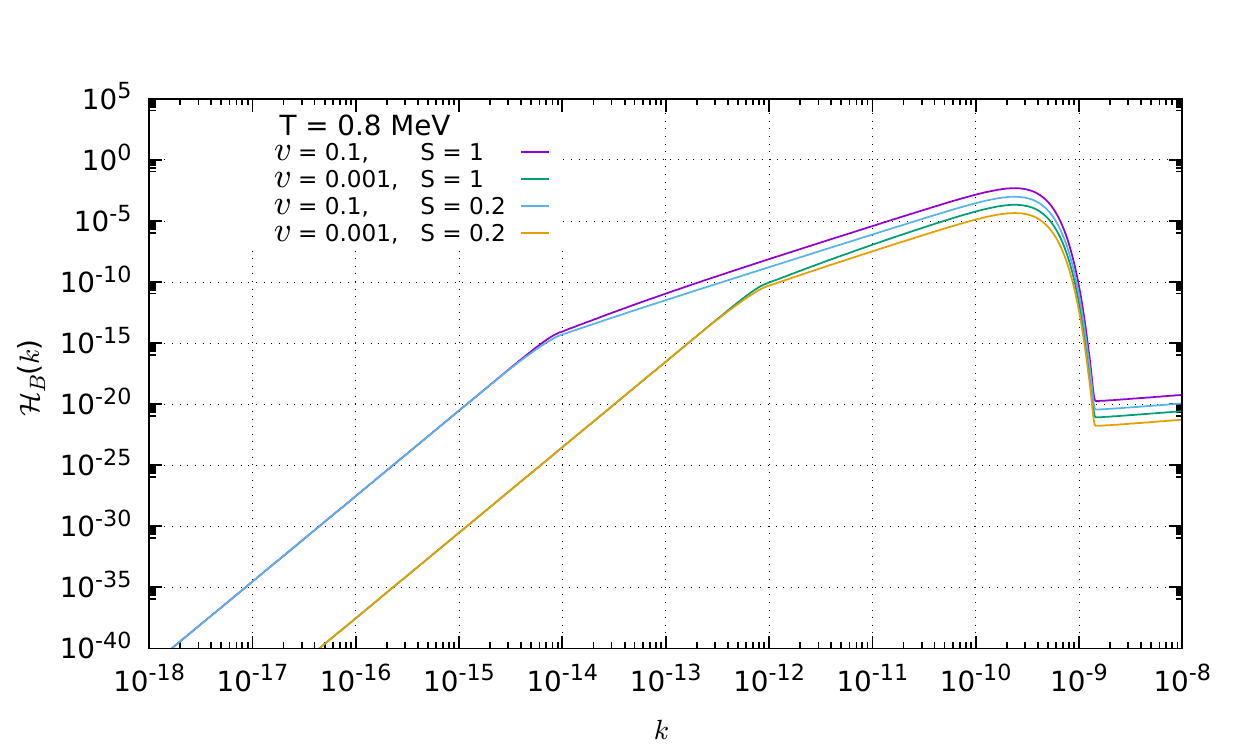}\label{fig:Hb}}
	\subfloat[]{\includegraphics[width=3.1in,height=2.2in]{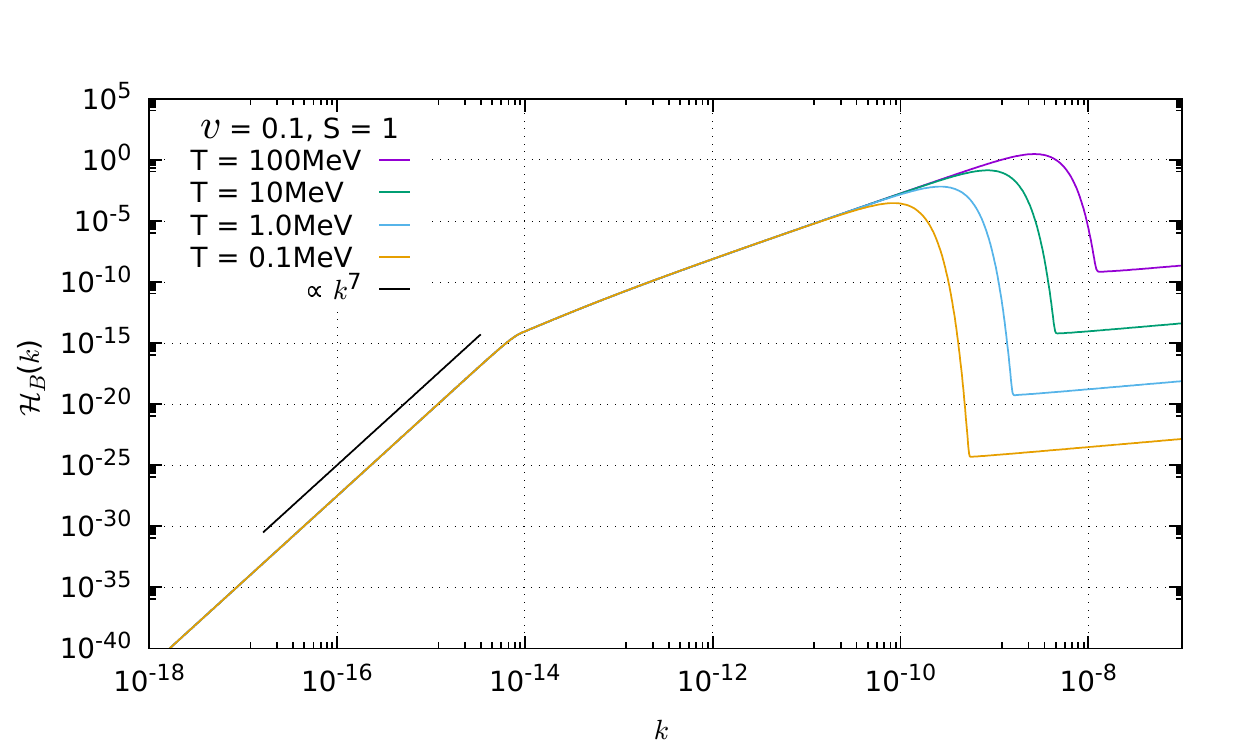}\label{fig:hb-2}} 
	\caption{ Plot of  helical energy with respect to wave number $k$ for fix $\sigma=70$. In figure (\ref{fig:Hb}), we keep fix temperature (0.8 Mev) and plot for different combinations of fudge factor (S) and $v_i(\tau_*)$. In figure (\ref{fig:hb-2}), we fix the fudge factor and $v_i(\tau_*)$, and plot for different temperatures.}
	\label{fig:Hb-tvs}
\end{figure*}
%
\section{Gravitational wave generation by the magnetic fields in a neutrino asymmetric plasma}\label{sec-3} 
In the present work, we have considered following perturbed metric $g_{\mu\nu} = a^2 (\tau)(\eta_{\mu\nu}+h_{\mu\nu}),$
%
%
%
where, $\eta_{\mu\nu} = \text{diag}(-1, 1, 1, 1)$, $h_{00} = h_{0i} = 0$, and $|h_{ij}|\ll 1 $. Here $h_{ij}$ represents the comoving tensor perturbation and satisfies the following gauge conditions $h_i^i=0$ and $\partial^i \, h^j_i=0$. This gauge choice in literature is known as Lorentz gauge or Newtonian gauge or TT gauge. In the linear perturbation theory, the metric perturbations in the TT gauge are gauge independent \cite{Bardeen:1980mj, Kodama:1984ms}. In this gauge, there are two independent polarization states of the tensor perturbation. The evolution equation for the tensor perturbation can be obtained by taking Einstein equation to the order $\mathcal{O}(h)$. The linear order of Einstein equation $\delta G_{ij}=8\pi G\, \delta T_{ij}$ \cite{Weinberg:100595} will give
\begin{equation}
-\frac{1}{2} h_{ij;\nu}^{;\nu} =8\pi G\, \Pi_{ij}^{TT} \label{eq:tensor-1}. 
\end{equation}
Here, $\Pi_{ij}^{TT}(t, {\bf x})$ is the anisotropic part of the stress tensor in TT gauge \eqref{eq:FLRWperturb}. Here semi-colons denotes the covariant derivative while comma denote  partial derivative. Under the general coordinate transformation, each component of the tensor perturbation may be treated as a scalar quantity i.e. $h_{ij;\mu}=h_{ij,\mu}$. Therefore, left hand side of the equation (\ref{eq:tensor-1}), gives
\begin{eqnarray}
h_{ij;\mu}^{;\mu} & = & \bar{g}^{\mu\nu} (h_{ij,\mu\nu}-\Gamma^\gamma_{\mu\nu}\, h_{ij, \gamma}) \nonumber \\
& = & - \ddot{h}_{ij}+\frac{\nabla^2}{a^2} h_{ij}- \frac{3 \dot{a}}{a} \dot{h}_{ij}\,,
\end{eqnarray}
here dot over the quantity represents derivative with respect to physical time $t$ and $\Gamma^0_{0\nu} =\Gamma^0_{0\mu}=0$, $\Gamma^0_{ij}=\frac{\delta_{ij}}{a^2}$. Hence, equation (\ref{eq:tensor-1}) in Fourier space is
\begin{equation}
\ddot{h}_{ij}(\tau, {\bf k}) + \frac{3\dot{a}}{a}\,\dot{h}_{ij}(\tau, {\bf k}) +\frac{k^2}{a^2}h_{ij}(\tau, {\bf k}) = \frac{16\pi G}{a^2}\, \Pi_{ij}^{TT}(\tau, {\bf k})
\label{eq:tensor-ptbn1}.
\end{equation}
The effect of the expansion of the Universe can be seen through the second term in the above equation. Converting the cosmic time derivative to the conformal time $\tau$, 
\begin{equation}
h^{''}_{ij}(\tau, {\bf k}) + \frac{2 a'}{a} \, h'_{ij}(\tau, {\bf k}) + k^2\, h_{ij}(\tau, {\bf k}) = 16\pi G\, \Pi_{ij}^{TT}(\tau, {\bf k})
\label{eq:tensor-ptbn}.
\end{equation}
Here $a'/a =1/\tau$ and $a^{'}/a =2/\tau$ are defined for radiation and matter dominated era respectively. This equation looks similar to the massless Klein Gordon equation for a plane waves in an expanding space-time with a source term. Since the source of the GWs is magnetic field generated via causal process and they are redshifted, we can write $ \Pi_{ij}^{TT}(\tau, {\bf k})= \Pi_{ij}^{TT}({\bf k})/a^2(\tau)$. 
The plane wave solution form is given by
\begin{eqnarray}
h_{ij}({\bf k}, \tau) =  \sum_{\alpha=+,-} \int \frac{d^3 k}{(2\pi)^3}\, h_\alpha ({\bf k}, \tau) e^{i{\bf k}\cdot {\bf x}} \varepsilon_{ij}^\alpha \, ,  
\end{eqnarray} 
where, we have chosen a orthonormal coordinate system ($\hat{e}_1$, $\hat{e}_2$, $\hat{e}_3= \hat{{\bf k}}$) in such a way that GW propagate in $\hat{e}_3$ direction and 
\begin{equation} \label{eq:tenpolari}
\varepsilon_{ij}^\pm =-\sqrt{\frac{3}{2}} \,(\varepsilon_i^\pm\times  \varepsilon_j^\pm), 
\end{equation}
satisfies the following relations: $\delta_{ij}\varepsilon^\pm_{ij}=0$, $\hat{k}_i \varepsilon^\pm_{ij}=0$ and $\varepsilon^\pm_{ij} \varepsilon^\mp_{ij}=3/2$. Using this basis defined in equation \eqref{eq:tenpolari}, we can decompose $\Pi_{ij}$ and $h_{ij}$ as
\begin{eqnarray}
\Pi_{ij}(k)  & = &  \Pi^+(k) \, \varepsilon_{ij}^+ + \, \Pi^-(k)\, \varepsilon_{ij}^- , \,\nonumber\\ 
h_{ij}({\bf k}, \tau) & = &  h^+({\bf k}, \tau)\, \varepsilon_{ij}^+ + h^-({\bf k}, \tau)\, \varepsilon_{ij}^-\,.  \nonumber
\end{eqnarray} 
Therefore, equation (\ref{eq:tensor-ptbn}) will reduced to
\begin{equation}
h^{\pm ''}({\bf k}, \tau)+ 2\,{a'\over a}\, h^{\pm '}({\bf k}, \tau)+k^2 h^{\pm}({\bf k}, \tau)=8\pi G {\Pi^{TT \pm}({\bf k})\over a^2}\, ,
\label{eq:hij-pm}
\end{equation}
To solve equation \eqref{eq:hij-pm}, we have defined a dimensionless variable $x=k\tau$. Then equation \eqref{eq:hij-pm} in a radiation dominated universe reduces to 
\begin{equation}
h^{\pm ''} + \frac{2}{x}\, h^{\pm '} + h^{\pm} =  \frac{S^\pm}{k^2}\, ,
\label{eq:GW-source}
\end{equation}
Here $S^\pm$ is the mean square root value of $\Pi^{TT \pm}$ and can be calculated by decomposing equation \eqref{eq:Pij-corre} into the polarization component. Mean square root value of the two stress energy momentum tensor polarization is given by
%
%
%
\begin{equation}
\langle|\Pi^{TT \pm}(k)| \rangle= \sqrt{\frac{f(k)\, \mp \, g(k)}{3}}\,.
\end{equation}
The solution of equation (\ref{eq:GW-source}) in radiation dominated era via  a relevant Green's function is
\begin{eqnarray}
h^{\pm}(x)= A^{\pm}(x)\frac{\sin x}{x}-B^{\pm}(x)\frac{\cos(x)}{x}\,,
\label{eq:hij-sol}
\end{eqnarray}
where the coefficients $A^\pm$ and $B^\pm$ is given by
\begin{eqnarray}
A^{\pm}(x) & = &  16\pi G \int_{x_{*}}^1 S^\pm(x') \frac{\cos(x')}{x'} \left(\frac{x'}{k}\right)^2 d x'\,, \\
B^\pm(x) & = &16\pi G  \int_{x_{*}}^1 S^\pm(x') \frac{\sin(x')}{x'} \left(\frac{x'}{k}\right)^2 d x'\,,
\end{eqnarray}
Here upper limit in the integral $x=1$ correspond to horizon size. Since second term diverges, for the modes of the horizon size i.e. $x\ll 1$, we will neglect second term. Therefore, solution of the equation in the present case is
\begin{alignat}{2}
h^+(x) & = &  A^+ \,\frac{\sin x}{x} \approx -2\left(\frac{M_*}{M_{\rm pl}}\right)^2\, \sqrt{\frac{f-g}{3}} \ln (x_{*})\, \frac{\sin x}{x} \label{h+x}\,,\\
h^-(x) & = &  A^-\, \frac{\sin x}{x} \approx -2\left(\frac{M_*}{M_{\rm pl}}\right)^2\, \sqrt{\frac{f+g}{3}} \ln (x_{*})\, \frac{\sin x}{x} \label{h-x}\,.
\end{alignat}
%
\subsection{Energy density of GW}
The (00) component of the stress-energy tensor is defined as the energy density of the gravitational waves and it is defined as
\begin{eqnarray}
\rho_{\rm GW}({\bf x}, \tau) =  
\frac{1}{16 \pi G \, a^2(\tau)}\langle h^{'}_{ij}({\bf x}, \tau)\, h_{ij}^{'*}({\bf x}, \tau) \rangle.
\end{eqnarray}
Here averaging is taken over the comoving volume of the horizon size as modes with $k\tau \ll 1$ cannot be defined in a meaningful way. Therefore energy density of the GWs in a Fourier space is written as:
\begin{eqnarray}
\rho_{\rm GW}= \int \frac{dk}{k}\, \frac{d \rho_{\rm GW}(k)}{d\, \ln(k)}\,,
\label{eq:energy-density-hij}
\end{eqnarray}
where,
\begin{equation}
\frac{d \rho_{\rm GW}(k)}{d\, \ln(k)} = \frac{k^3}{(2\pi)^6\, G\, a^2}[|h^{+'}|^2+|h^{-'}|^2]\,.
\end{equation}
The relative energy density $\Omega_{\rm GW}(\tau, {\bf x})$ of GW is given as 
\begin{eqnarray}
\Omega_{\rm GW}({\bf k}, \tau) \equiv \frac{\rho_{\rm GW}({\bf k}, \tau)}{\rho_{\rm cr}(\tau)}\,,
\end{eqnarray}
here, $\rho_{\rm cr}$ is the critical energy density of the Universe at a given time. The GW spectrum per logarithmic momentum interval at the time of generation is
\begin{eqnarray}\label{eq:power-1}
\left.\frac{\Omega_{\rm GW}}{d \ln {\bf k}}\right\vert_* = \left. \frac{1}{\rho_{\rm cr}}\frac{d \rho_{\rm GW}}{d \ln k}\right\vert_*=\frac{k^3}{(2\pi)^6\,\rho_{\rm cr}\, G\, a^2}[|h^{+'}|^2+|h^{-'}|^2].
\end{eqnarray} 
\subsection{Effective degree of freedom}
In radiation dominated era, the energy density of the Universe evolves as $\rho\propto a^{-4}$ and due to the interaction of particles with the photons leads to the thermal equilibrium. In an adiabatic system, the entropy per unit comoving volume should be conserved and therefore, $s(T) a^3(T)={\rm constant}$ (where $s(T)=\frac{\rho+p}{T}=\frac{2\pi^2}{45}g_{s}(T) T^3$). The energy density in this era is given by $\rho(T)=\frac{\pi^2}{30}g(T) T^4$ and $p=\frac{1}{3}\rho$. Here $g_{s}(T)$ and $g(T)$ are the effective degree of freedom of all relativistic species and are given as
\begin{eqnarray}
g=\sum_i g_{i}(T)\left(\frac{T_i}{T}\right)^4,   ~~~~~ 
g_{s}=\sum_i g_{s_i}(T)\left(\frac{T_i}{T}\right)^3.
\end{eqnarray}
Using the fact that the value of the quantity $[s(T) a^3(T)]$ at the time of generation and present time remain same, i.e., $[s(T) a^3(T)]_*=[s(T) a^3(T)]_0$,
\begin{equation}
\frac{a_*}{a_0}=\left(\frac{g_{s_0}}{g_{s_*}}\right)^{1/3}\frac{T_0}{T_*}\,.
\label{eq:edof-1}
\end{equation}
The present values of $g$ and $g_{s}$ are equal for temperature $T \gtrsim 0.1 MeV$ and are nearly equal for temperature $T \lesssim 0.1 MeV$ ($g_0= 3.3626$ and $g_{s_0}= 3.9091$).
%
\subsection{Observed energy spectrum of GW}
Once GWs are produced, they will propagate unhindered throughout the space. Which means, their energy density will fall only due to the expansion of the Universe and hence $\Omega_{\rm GW} \propto a^{-4}$. Hence the energy density of the GW at the time of generation in terms of today's value can be written as
\begin{eqnarray}
\left.\frac{d\Omega_{\rm GW}}{d \ln k}\right\vert_0 =\left.\frac{\Omega_{\rm GW}}{d \ln {k}}\right\vert_* \left(\frac{a_*}{a_0}\right)^4 \frac{\rho_{c,*}}{\rho_{c,0}}\,.
\label{eq:GWdensity-present-source}
\end{eqnarray}
Therefore, the observed power spectrum per logarithmic scale can be written using equations \eqref{eq:power-1} and \eqref{eq:edof-1} in \eqref{eq:GWdensity-present-source}
\begin{eqnarray} \label{eq:om*}
\left.\frac{d\Omega_{\rm GW}}{d \ln k}\right\vert_0  = \left.\frac{\Omega_{\rm GW}}{d \ln {k}}\right\vert_* \left(\frac{g_{s_0}}{g_{s_*}}\right)^{4/3}\left(\frac{T_0}{T_*}\right)^4 \frac{\rho_{c,*}}{\rho_{c,0}}\,.
\end{eqnarray}
Which can be simplified using $H^2=\frac{8\pi G}{3} \, \rho$ and equations \eqref{eq:power-1}, \eqref{h+x} and \eqref{h-x} in equation \eqref{eq:om*}
\begin{eqnarray}
\left.\frac{d\Omega_{\rm GW}}{d \ln k}\right\vert_0 & = &  \frac{64\pi k^3}{9(2\pi)^6}\, \left(\frac{90}{8\pi^3 g_{s*}}\right)^2 \left(\frac{g_{s0}}{g_s*}\right)^{4/3} \, \left(\frac{T_0}{T_*}\right)^2\, \left(\frac{T_0}{H_0}\right)^2 \nonumber\\
&\times & \left[\ln(x_{*})\frac{\partial }{\partial \tau}\,\left.\left(\frac{\sin x}{x}\right)\right\vert_{\tau=\tau_*}\right]^2 \, f(k)\,,
\end{eqnarray}
here, $x_{\rm in}=k_{*} \tau_*$ and $x = k\tau$. $H_0=2.133\times 10^{-42}$ GeV, $g_{s0}=3.7$ value at $T_0 = 2.23\times10^{-13}$ GeV, $g_{s*}$ at $T_*$ i.e. temperature of the magnetic field generations. In above equation 
\begin{eqnarray}
\frac{\partial }{\partial \tau}\left.\left(\frac{\sin x}{x}\right)\right\vert_{\tau=\tau_*} & = & \frac{1}{k\tau_*}[k\tau_* \cos(k\tau_*)- \sin(k\tau_*)] \nonumber\,, \\
\ln(x_*) & = & \ln(k_*\tau_*) \nonumber\,.
\end{eqnarray}
\begin{figure}[ht]
    \includegraphics[width=3.2in,height=2.2in]{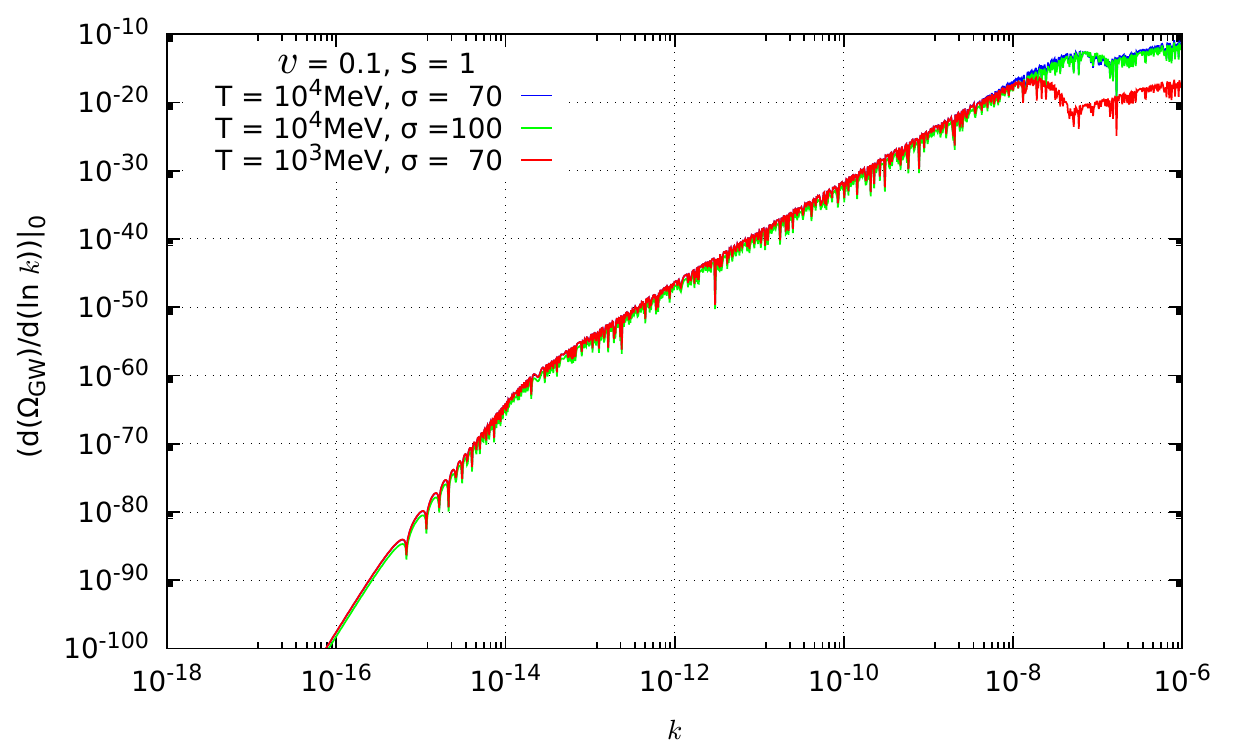}\label{gww1d4m}
	\caption{GW power spectrum with respect to a dimensionless variable $k$ at different temperatures for a fix fudge factor (S = 1) and $v_i(\tau_*)=10^{-1}$ . The peak corresponds to the frequency $\nu\sim 10^{-10}$ Hz. Since the amplitude $\sim 10^{-12}$ and frequency $\sim 10^{-10}$ Hz comes under the sensitivity of the SKA and IPTA, the generated GWs may be detected in these observations. The plot also shows how the power spectrum vary with comoving electric conductivity.}
	\label{fig:Gw-power}
\end{figure}
\section{Results and Discussions}\label{sec-4}
In the present work, we have studied the generation of magnetic fields and hence the GWs in a hot dense plasma for the case of $T\gg \mu$ (all plots are obtained for the this case). In this plasma, a net neutrino number densities
are present. In the presence of this net neutrino number density, turbulence is generated and hence magnetic fields are produced. It is to be noted here that, we have not considered any fluid viscosity in the plasma, which normally act at very small length scale or at large $k$ values. The value of $k$, for which viscosity works, is greater than $k_{\rm diss}\sim \sqrt{\frac{g_{\rm eff}}{6}}\alpha_{\rm em}T$ \cite{Weldon:1982ha}. Therefore, the growing mode at large $k$, in our case, is not physical and may not be of importance. At these length scale, dissipative effects may become important \cite{Brandenburg:1996aek}. In figures (\ref{fig:detail}) and (\ref{fig:Hb-tvs}) we have shown the evolution of magnetic energy, turbulent kinetic energy, and the helical energy for different parameters. We have shown in figure (\ref{fig:Eb-plot}) that the peak of the magnetic energy shifted towards lower $k$ values as we decrease the temperature. Initially, when there are no magnetic fields present, the last term in the equation dominates and hence the behavior of $k^8$. However, when sufficiently large magnetic fields are generated, second and third term start contributing. For a maximally helical magnetic field, magnetic modes grow exponentially for $k\sim \Sigma_2/2$, which can be seen from the first term in the equation \eqref{eq:EB-bv2}. We have also shown in figure \eqref{fig:Ev-plot} that at large values of the $k$, turbulent kinetic energy follows, Kolmogorov spectrum. However, for smaller $k$ values, the spectrum is a white noise spectrum. The variation of the helical energy density of the magnetic fields is given in plot \eqref{fig:Hb-tvs}. In figure \eqref{fig:Hb}, helical energy for different values of the Fudge factor are shown. It is clear that at a fixed temperature, peak remains at the same position. However, at a smaller value of $k$, power in the magnetic fields depends on the values of $v_i$ but not on the values of the fudge factor. It is interesting to note that peak shift at a smaller value of $k$ in figure \eqref{fig:hb-2}, which means that magnetic power shifts from small length scale to large length scale. This phenomenon is known as inverse cascading of magnetic energy. It is also shown here that slop at small $k$, is proportional to $k^7$. In reference \cite{Dvornikov:2015lea}, authors have shown that at neutron star core, a seed magnetic field of the order of $10^{12}$ G, modifies to $10^{17}$ G at a time scale of ($10^{3}-10^{5}$ yr). In this work, authors have shown that at core there are two contribution to the electric current: one from the chiral asymmetry and second one is from the non-zero weak interaction proportional to the finite neutron and proton densities. In Ref. \cite{Dolgov:2001dd}, it is shown that the magnetic modes grow exponentially by the electron neutrino asymmetry for the $\nu$ burst of a supernova explosion. Authors have argued that the generated magnetic fields via this mechanism may explain the strongest magnetic fields in magnetars. However, in the present work, we have shown that, we don't need a seed field and the magnetic fields are generated at the cost of turbulent energy in presence of neutrino asymmetry. Here one assumption is that net lepton density is homogeneous. We estimate the strength of the magnetic fields, which is $B\sim 4\pi \alpha \Sigma_\omega/(Tl_B)$. At temperature $T\sim$ MeV and length scale $l_B\sim $ MeV$^{-1}$, a seed magnetic field of the order of $10^{14}$ G could be generated which after amplification via a dynamo mechanism can give currently observed magnetic fields. These generated magnetic fields would create the anisotropy in the energy-momentum tensor, which in turn generate GW background. 
The power spectrum of the GW is given in plot  \eqref{fig:Gw-power} with respect to $k$ for the parameters given below the plot. 
The green plot represent the case when $\sigma=100 $ when $\sigma$ is reduced to 30\% , as shown by the blue curve, the peak-position and amplitude of the power-spectrum does not change significantly.
 At a given value of $\sigma$ and different temperatures, peak shifts towards lower $k$ value. It means that the inverse cascading behavior of the helical magnetic fields reflects here. At temperature $T=10^3$ MeV and $T=10^4$ MeV amplitude is $10^{-12}$ and $10^{-15}$ and corresponding to frequencies $10^{-10}$ Hz and $10^{-12}$ Hz respectively. Since the amplitude and frequency of the produced GWs lies in the sensitivity of the SKA and PTA, these produced GWs may be detected in these observations. 
%
Next, we compare our results with those obtained in Refs. \cite{Dolgov:2002gn, Dolgov:2001dd}. In these papers, authors have considered inhomogeneity in one or more neutrino species. Their results show that the primordial gravitational waves has a amplitude of the order of $h_0^2\Omega_{\rm GW}(f)\sim 10^{-12}$ with frequency $1$ mHz  at temperature around $100$ GeV, which may be detected in eLISA \cite{Dolgov:2002gn}. 

In summary, we have studied the generation of the magnetic and GWs in a $\nu \bar{\nu}$ gas in a hot dense plasma where there is a homogeneous net neutrino number densities are present. We predicts that the produced GW can be detected in SKA or PTA observations as the amplitude and the frequency lies in the sensitivity range of these observations. 
\section*{Acknowledgment} 
A.K.P. is financially supported by the Dr. D.S. Kothari Post-Doctoral Fellowship, under the Grant No. DSKPDF Ref. No. F.4-2/2006 (BSR)/PH /18-19/0070. A.K.P. also likes to thanks facility provided at IRC (IUCAA Resource Center), Department of Physics and Astrophysics, University of Delhi, India.
\bibliographystyle{apsrev4}
\bibliography{ref-neutrino}

\end{document}